# Dissociative excitation of nitromethane induced by electron impact in the ultraviolet – visible spectrum

Juraj Országh[1], Anita Ribar[1], Marián Danko[1], Dennis Bodewits[2], Štefan Matejčík[1], Wiesława Barszczewska[3]

[1]Department of Experimental Physics, Comenius University in Bratislava, Mlynská dolina F2, 842 48 Bratislava, Slovakia

[2]Physics Department, Leach Science Center, Auburn University, Auburn, AL 36832, USA

[3]Faculty of Exact and Natural Sciences, Siedlce University, 3 Maja 54, Siedlce 08-110, Poland

The excitation of nitromethane ($CH_3NO_2$), which is an important propellant and prototypic molecule for large class of explosives, has been investigated by electron impact and subsequent emission of photons in the UV-VIS spectral region between 300 nm and 670 nm. Emission spectrum of nitromethane was recorded at an electron energy of 50 eV. New dissociative excitation channels were discovered through the appearance of different CH, CN, NH, OH and NO bands, and the Balmer series of atomic hydrogen. In addition, relative emission cross sections were recorded for the transitions of selected fragments. The emission spectrum was captured at significantly higher resolution in comparison to previous studies.

## 1 Introduction

Nitromethane ($CH_3NO_2$) and other nitroalkanes are essential compounds in various explosives and propellants. Nitrogen oxides are among the gaseous products of detonation, so called blasting gases. These are toxic and in some applications of explosives, such as mining, they can cause increased health risks [1]. Explosion residues detection techniques are important to investigation the cause of explosions and are based on optical detection methods, especially in the mid-infrared spectral region [2-3], or on chemiluminescence [4-5]. The interaction of electrons and photons with nitromethane in the gas phase is relatively complex and has been investigated in the past [6-13]. The production of negative ions using dissociative electron attachment to nitromethane has been studied using the mass spectroscopy for analysis of the fragment negative ions [6,7]. Elastic scattering of the nitromethane has revealed the electronic structure of the molecule [9-11] and dissociative electron attachment (DEA) provided insight into resonant processes of nitromethane and low energy (0 – 16 eV) impact electrons [7]. However, there are few studies reporting electron excitation and dissociative excitation of nitromethane. Particularly, there are no recorded electron induced emission spectra of nitromethane at the ultraviolet-visible wavelengths. Photon induced fluorescence of the compound has been studied in the past by Butler et al. [8], and partially served as a benchmark for this work. In their work, experimental data on photodissociation of nitromethane using 193 nm laser beam were shown for both single-photon excitation and multiphoton excitation. Electron impact provides several experimental advantages over studies using photons. It can induce excitation transitions with change of multiplicity. Electron source also provides a variable energy in larger range which, in case of photoexcitation, is possible only by using synchrotron. That makes it possible to detect the interaction of electrons with nitromethane near the activation threshold of different reaction channels, as well as the saturation limit of the fluorescence originating from excited neutral species formed in the dissociative process. The emission spectrum of nitromethane with the discussion of the main features, as well as relative emission cross sections of several neutral fragments, such as H, CH, CN, NH, and NO bands are shown in this paper.

## 2 Experimental setup

The measurements were performed on the electron induced fluorescence apparatus at Comenius University Bratislava, previously described in [14, 15]. In short, the device (Figure 1) consists of a vacuum chamber with background pressures typically below $10^{-8}$ mbar. In the reaction chamber, an effusive molecular beam formed by a capillary is crossed perpendicularly with a beam of electrons, produced by an electron gun (EG). The electrons interact with the effusive molecular beam while the background pressure in the vacuum chamber is $10^{-4}$ mbar still ensuring binary electron-molecule collisions. The vacuum chamber background pressure, with the gas inlet closed, is $10^{-8}$ mbar. Electrons are thermally emitted from a tungsten hairpin filament, *Agar Scientific A054*. To avoid background photon radiation of the filament it is mounted in the cavity of the first and the second electrodes, which are covered by colloidal graphite. Electron currents are typically in the range of μA with an energy resolution of 0.8 eV FWHM. The kinetic energy can be varied in range of approximately 0 eV to 100 eV.

This study was demanding from the experimental point of view due to fast deterioration of the electron gun hot filament by nitromethane. The usual lifetime of the filament is several weeks or months, depending on the nature of studied compound it is exposed to (the setup is not differentially pumped). In the case of nitromethane, its lifetime was limited to approximately a day, most probably due to oxidation [16]. This caused problems, as each exchange of the filament and subsequent time for achieving of proper working conditions can take more than one day as well. For this reason, the nitromethane study was interrupted many times. The electron induced fluorescence studies are also very time consuming due to low photon production rates. To achieve better optical resolution and better signal-to-noise ratio, we accumulate signal for long time (up to several days). In present case it was not possible, and we had to compromise between the optical resolution and the accumulation time.

The electron energy was calibrated by measurement of the cross section energy dependence of the (0,0) band of $N_2$ (C $^3\Pi_u$ – B $^3\Pi_g$) at 337 nm [14] and He I (1s4d $^3D_{1,2,3}$ – 1s2p $^3P^0_{1,2}$) 447.14 nm emission line. The $N_2$ (C $^3\Pi_u$ – B $^3\Pi_g$)(0,0) cross section exhibits a pronounced peak at 14.1 eV [17] and the He I (1s4d $^3D_{1,2,3}$ – 1s2p $^3P^0_{1,2}$) cross section transition exhibits a step increase at 23.736 eV [18].

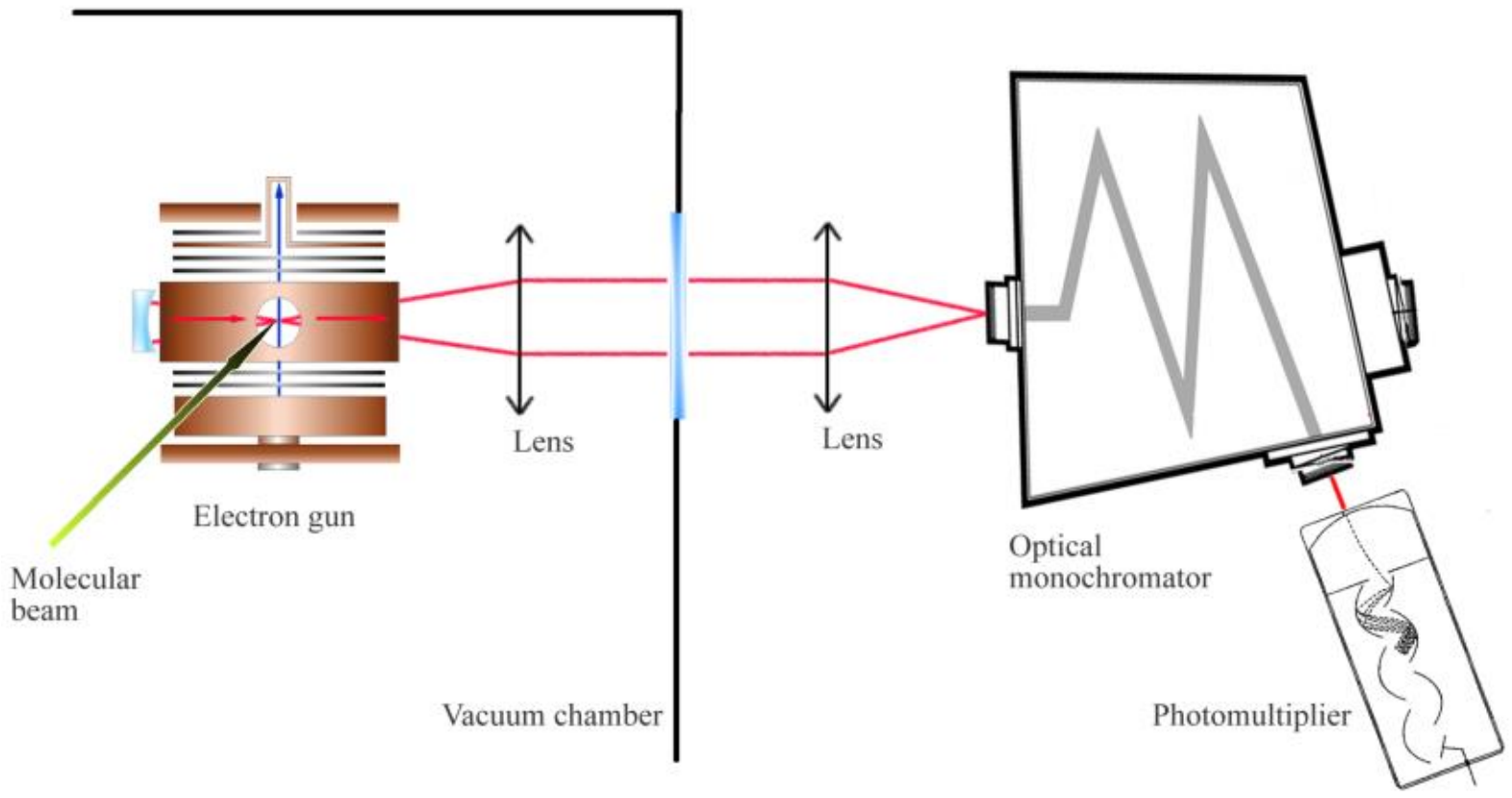

**Figure 1. Schematic outline of the electron impact apparatus at Comenius University Bratislava [15].**

Collection and detection of the optical signal is done by optical elements (mirror, lenses, window, and optical monochromator) and a photomultiplier (see Figure 1.). For our study of nitromethane, we used the *Oriel Cornerstone*$^{TM}$ *260 Czerny – Turner* optical monochromator. Its optical resolution is determined by the width of the entrance and exit slits. At an opening of 200 μm , we measured the width of the Ar 294.3 nm line to be approximately 0.7 nm FWHM ; with an 100 μm opening a resolution of approximately 0.4 nm resolution was achieved. The lenses directing the beam onto the entrance slit are made of UV-fused silica, the mirror of UV-enhanced aluminum, and the window is made of $MgF_2$. A thermoelectrically cooled photomultiplier (*Hamamatsu R4220P),* working in the photon counting regime, was used to detect the emission passing through the optical monochromator. The relative intensity of the recorded emission spectrum was calibrated to account for to the wavelength-dependent sensitivity of the optical system.

# 3 Experimental results and discussion

## 3.1 Emission spectrum

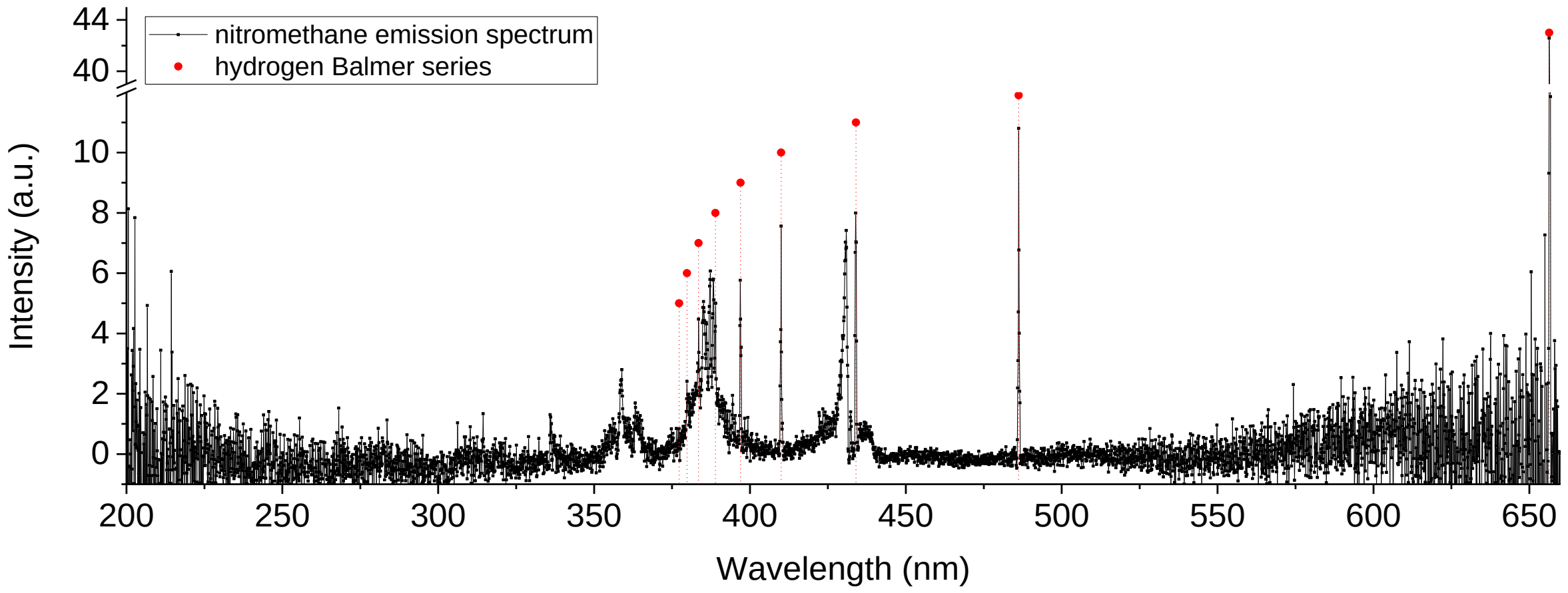

**Figure 2. The overview spectrum obtained by electron impact of $CH_3NO_2$ at an incident energy of 50 eV , calibrated to the apparatus' sensitivity. The H-Balmer series lines are marked by red dots.**

An overview of the resulting emission spectrum due to electron bombardment of nitromethane is shown in Figure 2. It was recorded in the spectral region 200 – 670 nm at an electron impact energy of 50 eV. The optical resolution of the spectrometer in the Figure 2 was approximately 0.4 nm throughout the bandpass (resolving power ~ 1000, allowing us to resolve various features of the spectrum. However, the signal intensity at the detector was low. The lines corresponding to the hydrogen Balmer series are marked by red dots and their positions are summarized in Table 1. We were able to identify Balmer series lines up to $H_l$ (11 – 2). The transitions from higher levels were too weak, blended with the background noise, and were therefore not detected in our spectrum. The lifetime of hydrogen states increases with increasing excitation level, up to several microseconds for the $H_d$ and above (Table 1, [15]). That allows a significant portion of excited atoms to leave the apparatus' field of view before emitting photon, further decreasing the intensity of the lines associated with transitions higher levels.

**Table 1.** Hydrogen Balmer series lines observed in the electron fluorescence spectrum of nitromethane at 50 eV electron impact energy. The reported wavelengths are in air and the lifetimes were calculated from Einstein coefficients.

| *Hydrogen line* | *Level (initial – final)* | *Wavelength (nm)* | *Lifetime (ns)* |
|---|---|---|---|
| $H_\alpha$ | 3 – 2 | 656.3 | 22.7 |
| $H_\beta$ | 4 – 2 | 486.1 | 119 |
| $H_\gamma$ | 5 – 2 | 434.1 | 395 |
| $H_\delta$ | 6 – 2 | 410.2 | 1030 |
| $H_\varepsilon$ | 7 – 2 | 397.0 | 2280 |
| $H_\zeta$ | 8 – 2 | 388.9 | 4520 |
| $H_\eta$ | 9 – 2 | 383.5 | 8226 |
| $H_\theta$ | 10 – 2 | 379.9 | 14040 |
| $H_\iota$ | 11 – 2 | 377.2 | 22742 |

The emission spectrum consists of atomic emission lines originating from neutral hydrogen atoms (Figure 2) and band structures (expanded views are shown in Figures 3, 5 and 6) from molecular fragments, both of which are formed by dissociative excitation (DE) and dissociative ionization and excitation (DIE) of the nitromethane induced by electron impact. Below we will discuss the emission from the individual fragments.

### 3.1.1 Atomic fragments

All the atomic lines identified in the spectrum correspond to the Balmer series of atomic Hydrogen (Table 1). Interestingly, no other atomic emission lines were detected in our spectra. For example, in the spectrum of acetylene ($C_2H_2$) measured at the same electron impact energy of 50 eV, both C I and C II lines are clearly visible [15].

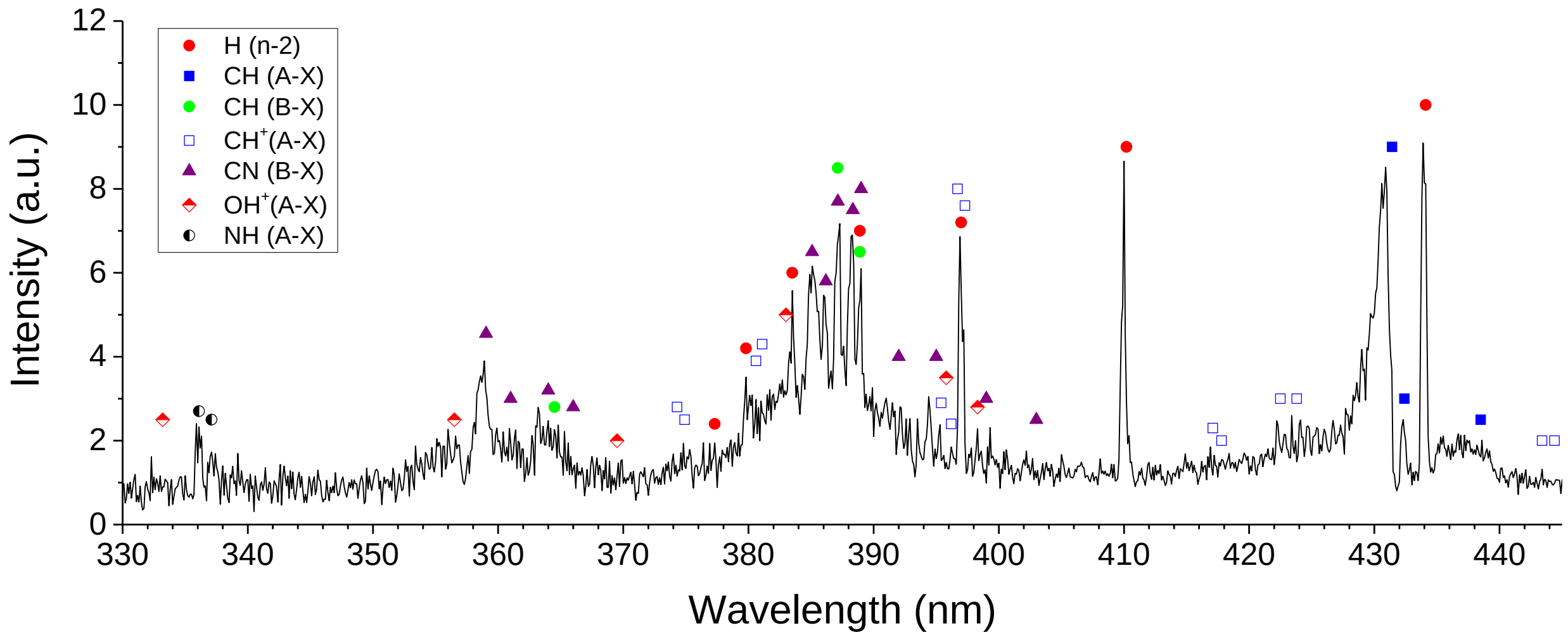

**Figure 3.** Most prominent molecular bands and Balmer lines in nitromethane emission spectrum obtained at 50 eV and 0.4 nm optical resolution identified according to [15] and [19-23]. The spectrum has been corrected for the wavelength dependent sensitivity of the apparatus.

### 3.1.2. Methylidyne (CH)

The second dominant feature of the resulting spectrum at 50 eV electron impact is attributed to fluorescence emission from transitions of the CH radical and its ion. The envelope of the rotational structure of CH (A $^2\Delta \rightarrow$X $^2\Pi$) (0, 0) (beginning with P branch head at 438.5 nm through Q branch band head at 431.4 nm) is visible in Figure 3, with the tail overlapping weaker $CH^+$ (A $^1\Pi \rightarrow$ X $^1\Sigma$) (0, 0) Q and R branches. Another prominent CH feature lays in the violet region. It is composed of CH (B $^2\Sigma^- \rightarrow$X $^2\Pi$) vibrational transitions – the band head of (0,0) Q branch at 388.9 nm is detected between the emission lines of Balmer series, $H_\zeta$ and $H_\eta$ at 388.9 nm and 383.5 nm, respectively. Strong Q branches of CH (C $^2\Sigma^+ \rightarrow$ X $^2\Pi$) (0, 0) at 314.5 nm (1, 1) at 315.7 nm contribute to the photon emission signal.

### 3.1.3 Cyanogen (CN)

The CN (B $^2\Sigma^+ \rightarrow$ X $^2\Pi$) band strongly overlaps with the CH (B $^2\Sigma^- \rightarrow$X $^2\Pi$) band between 383 nm – 389 nm and between 362 nm – 368 nm (where higher Balmer series lines can contribute as well), thus further investigation is needed in order to determine relative intensity contribution of the two to the overall signal. To further analyze these results the theoretical CN (B $^2\Sigma^+ \rightarrow$ X $^2\Pi$) spectrum was calculated using Lifbase software [23] (Figure 4). An interesting feature is present in this part of the spectrum: many of the CN (B $^2\Sigma^+ \rightarrow$ X $^2\Pi$) tail bands belong to a structure that is clearly observable despite its relatively low intensity. These bands are reported by Pearse and Gaydon [20] as occurring when the CN (B $^2\Sigma^+ \rightarrow$ X $^2\Pi$) system is strongly developed (as in active nitrogen or carbon arcs).

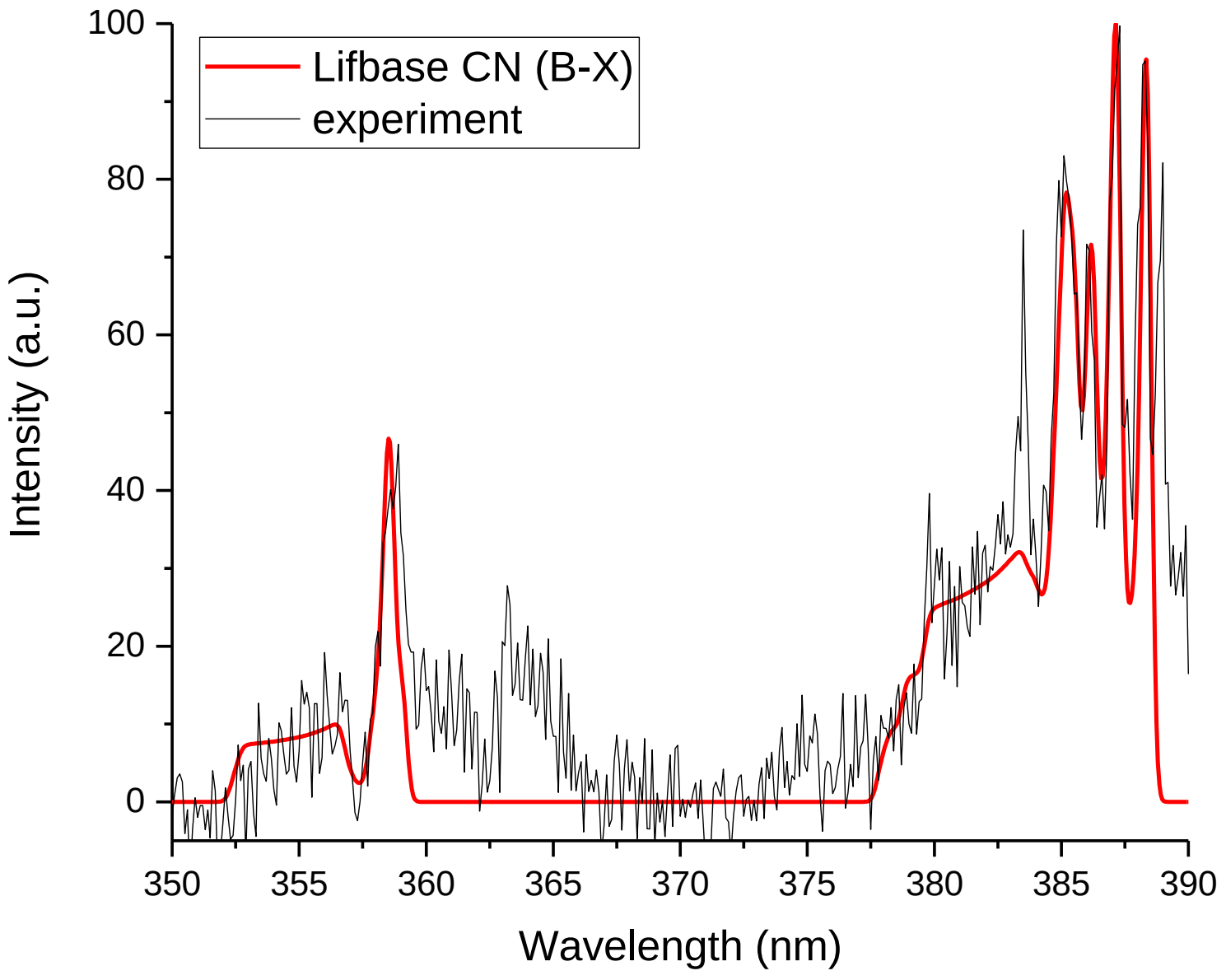

**Figure 4. Comparison** of experimentally obtained spectrum of the CN (B-X) band with a model calculated using Lifbase [23].

The tail bands in the sections 352 – 357 nm and 377.7 – 384 nm exist due to population of higher rotational states. The Lifbase calculated CN (B $^2\Sigma^+ \rightarrow$ X $^2\Pi$) spectrum, with a constant distribution of rotational states between levels N = 10 and N = 66 for each vibrational level, was found to approximately follow the shape of the experimental spectrum. The relative distribution of vibrational states in the theoretical spectrum is shown in Table 2.

**Table 2. Relative population of vibrational states of CN (B $^2\Sigma^+$) according to Lifbase simulation.**

| Vibrational level v` | Relative contribution |
|---|---|
| 0 | 0.310 |
| 1 | 0.245 |
| 2 | 0.146 |
| 3 | 0.137 |
| 4 | 0.162 |

## 3.1.4 Imidogen (NH)

Further relatively intensive band detected corresponds to the transition of NH (A $^3\Pi \rightarrow$X $^3\Sigma^-$) between 335.5 nm and 345.5 nm. The Q branch band heads (0, 0) and (1, 1) are clearly visible and the vibration modes of the P branch in the range 337.5 nm – 345.5 nm are very weak and more or less blended in with the background noise.

**Table 3. Diatomic emission bands originating from dissociative excitation of nitromethane at 50 eV electron impact energy. The reported wavelengths are in air.**

| Species | Electron transition | Vibr. transition (v`, v``) | Branch | Wavelength [nm] | Reference |
|---|---|---|---|---|---|
| OH | A $^2\Sigma^+ \rightarrow$ X $^2\Pi$ | (0, 0) | R | 306.3 | [20] [21] |
| | | | Q | 308.9 | [20] [21] |
| | | (1, 1) | R | 312.2 | [20] [21] |
| $OH^+$ | A $^3\Pi \rightarrow$ X $^3\Sigma$ | (0, 0) | | 356.5 | [20] [21] |
| CN | B $^2\Sigma^+ \rightarrow$ X $^2\Pi$ | (0, 0) | | 388.3 | [22] [19] |
| | | (1, 1) | | 386.9 | [19] |
| | | (2, 2) | | 386.0 | [19] |
| CH | A $^2\Delta \rightarrow$X $^2\Pi$ | (0, 0) | Q | 431.4 | [20] [22] |
| | B $^2\Sigma^- \rightarrow$X $^2\Pi$ | (0, 0) | R | 387.1 | [20] [22] |
| | | | Q | 388.9 | [20] [22] |
| | C $^2\Sigma^+ \rightarrow$ X $^2\Pi$ | (0, 0) | Q | 314.5 | [20] |
| | | (1, 1) | Q | 315.7 | [20] |
| $CH^+$ | A $^1\Pi \rightarrow$ X $^1\Sigma$ | (0, 0) | Q | 422.5 | [20] |
| NH | A $^3\Pi \rightarrow$X $^3\Sigma^-$ | (0, 0) | Q | 336.0 | [19] |
| | | (1, 1) | Q | 337.1 | [19] |
| NO | A $^2\Sigma^+ \rightarrow$X $^2\Pi$ | (0, 1) | | 239.8 | [20] [23] |
| | | (0, 2) | | 251.0 | [20] [23] |
| | | (0, 3) | | 263.5 | [20] [23] |
| | | (0, 4) | | 276.0 | [20] [23] |
| | | (0, 5) | | 289.5 | [20] [23] |
| | | (1, 4) | | 303.9 | [20] [23] |
| | B $^2\Pi \rightarrow$X $^2\Pi$ | (3, 3) | | 233.1 | [20] [23] |
| | | (1, 2) | | 233.9 | [20] [23] |
| | | (2, 3) | | 238.3 | [20] [23] |

| | | |
|---|---|---|
| (0, 2) | 239.8 | [20] [23] |
| (3, 4) | 243.0 | [20] [23] |
| (1, 3) | 244.4 | [20] [23] |
| (2, 4) | 249.6 | [20] [23] |
| (0, 3) | 251.0 | [20] [23] |
| (3, 5) | 254.1 | [20] [23] |
| (1, 4) | 256.0 | [20] [23] |
| (0, 4) | 263.5 | [20] [23] |
| (2, 6) | 273.4 | [20] [23] |
| (0, 5) | 276.0 | [20] [23] |
| (1, 6) | 281.1 | [20] [23] |
| (0, 6) | 289.5 | [20] [23] |
| (3, 8) | 292.0 | [20] [23] |
| (1, 7) | 294.5 | [20] [23] |
| (0, 7) | 303.9 | [20] [23] |

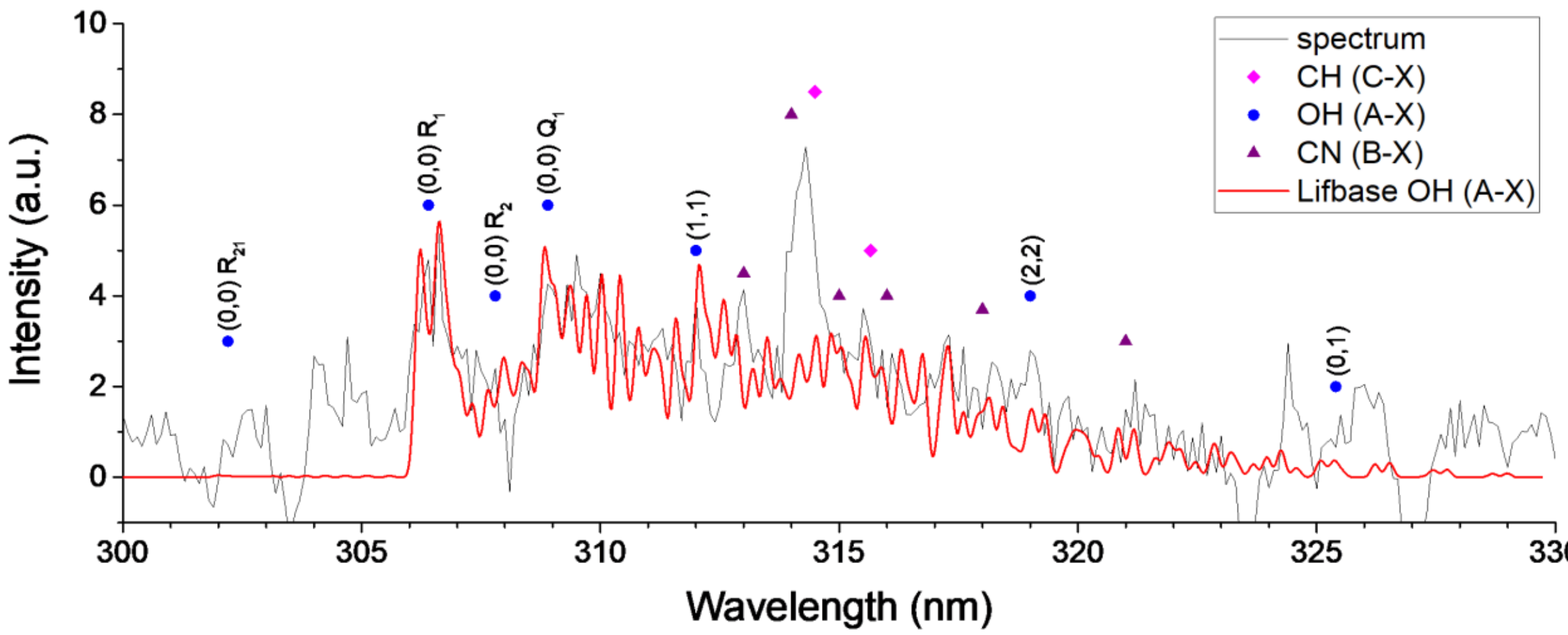

**Figure 5. OH radical emission overlapping with CH and CN band emission obtained at lower resolution (0.8 nm).**

### 3.1.5 Hydroxyl (OH)

The resulting spectrum between 300 nm – 330 nm is presented in Figure 5. Due to the low signal-to-noise ratio, we decreased the spectral resolution to 0.8 nm, resulting in higher photon yields, as compared to the spectral resolution of 0.4 nm which was used to obtain the spectra shown in Figures 2 and 3. A weak signal of the OH violet band OH (A $^2\Sigma^+ \rightarrow$ X $^2\Pi$) in the spectrum is assigned to rotational band heads of the (0, 0) and (1, 1) transitions (Table 3). Since the OH features are overlapping with the CH and CN bands in the spectrum and due to low SNR it is difficult to identify individual OH (A $^2\Sigma^+ \rightarrow$ X $^2\Pi$) bands. We again used Lifbase [23] to reproduce the experimentally acquired. The simulated spectrum contains vibrational levels v` = 0 and v` = 1 in the ratio 56:44 and for simplification the rotational population of each level was set to a thermal distribution corresponding to 8000 K. In case of electron induced excitation, rotational populations are usually not thermal; therefore, there are visible discrepancies between simulated and experimental spectrum. However, the overall shape of the measured and simulated bands is similar. These observations also suggest higher internal energy of OH (A $^2\Sigma^+$) similarly to other excited products of electron impact on nitromethane.

Even though there is no direct bond between the oxygen and hydrogen atoms in nitromethane molecule the generation of OH by electron impact was previously observed in a mass spectrometric study, focused on the formation of negative ions [7]. The authors suggested that $OH^-$ generation can be caused by electron reaction with vibrationally excited nitromethane. According to the calculations of thermal rearrangement of nitromethane by McKee [24] the rearrangement to aci-nitromethane $CH_2N(O)OH$ containing the O-H bond is possible at energies as low as 75 kcal/mol, which is approximately 3.25 eV per molecule. Hence, 50 eV electron impact could cause such rearrangement and further fragmentation and excitation of OH radical as well.

### 3.1.6 Nitric Oxide (NO)

The section of the spectrum between 230 nm and 300 nm contains various vibrational transitions of the nitric oxide NO (B $^2\Pi \rightarrow$ X $^2\Pi$) - β system with some weaker transitions of NO (A $^2\Sigma^+ \rightarrow$ X $^2\Pi$) - γ system (Figure 6). These were identified according to [20] and [23] and are summarized in the Table 3.

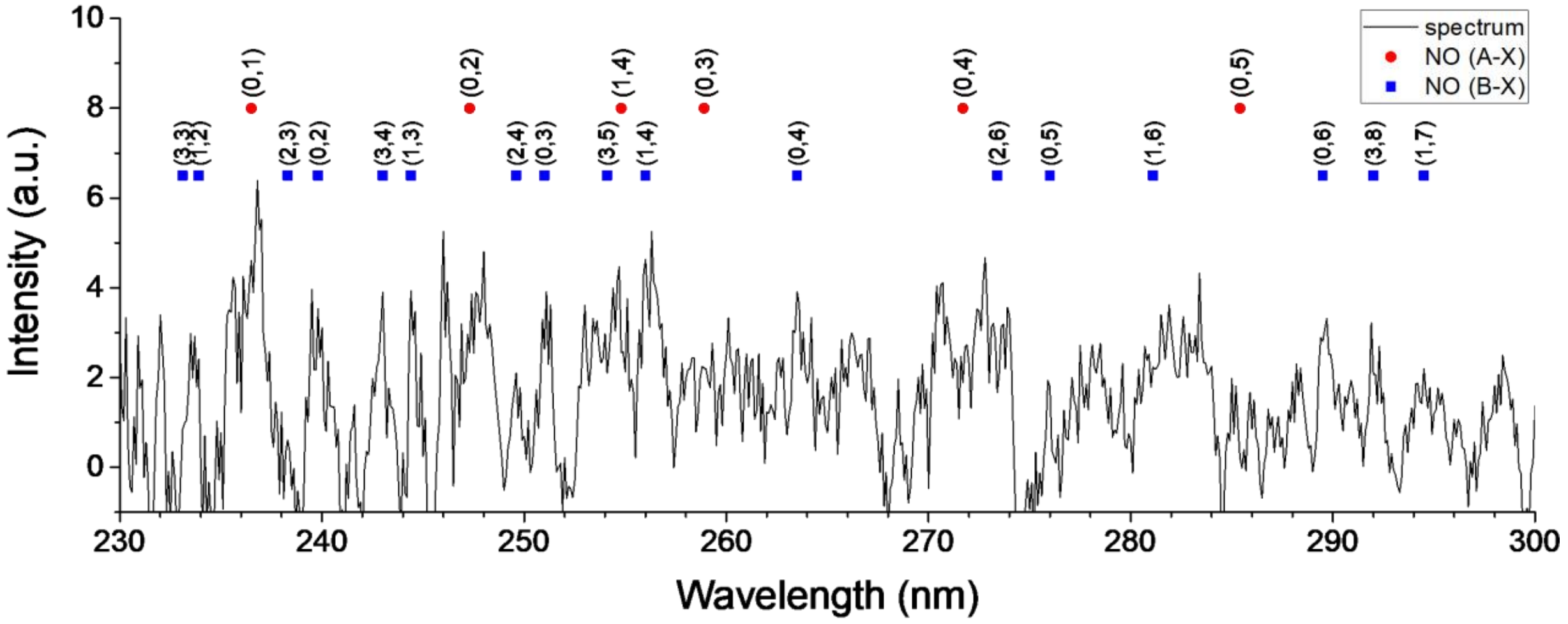

**Figure 6. NO transitions (A-X) and (B-X) form a relatively dense spectrum in the UV region between 230 nm and 300 nm. The spectrum was obtained at lower resolution (0.8 nm).**

### 3.1.7 Nitrogen Dioxide ($NO_2$)

Surprisingly, we detected no evidence of continuum emission from $NO_2$, which is one of the strongest emission features in other experiments of the fluorescence of nitromethane, cf. Butler et al. [8]. They obtained a spectrum in the region between 400 nm and 800 nm at pressures of 20 mTorr by photolysis of nitromethane using a 193 nm excimer laser (single photon excitation). The second measurement of nitromethane done by the same group was performed using a focused laser beam (multiphoton excitation). The most significant feature of the two spectra, as shown in Figure 7, together with our electron induced spectrum, is the strong continual signal rising above 398 nm. According to Butler et al. [8], this continual radiation corresponds to $NO_2$ continuum. There are two major differences between the two spectra published by Butler et al.: in the focused laser beam spectrum the maximum of the $NO_2$ dispersed fluorescence is shifted to lower wavelengths (higher energies), and the CH and CN bands are superimposed on $NO_2$ emission. The latter feature is an indication of multiphoton processes. Both dispersed spectra reported by Butler et al. as shown in Figure 7 are normalized to the same level of maximum fluorescence intensity for comparison. The intensities of the spectra published in [8] are not corrected for the sensitivity of their detectors, and therefore only the shape of the emission can be compared directly. The broad radiation of $NO_2$ is caused by its transition from the excited state to dissociative state leading to formation of NO and O. Butler et al. reported the emission from 398 nm, which is the dissociation limit of $NO_2$ to NO + O. The lack of discrete structures in the continuum emission of the $NO_2$ was attributed to strong vibronic coupling between the X $^2A_1$ ground state and the A $^2B_2$ state [25]. The absence of the continuum in the electron-induced experiment suggests that other dissociation channels are preferred and $NO_2$ is not formed, that the $NO_2$ does not have enough internal energy to undergo the dissociation process to NO + O, or most probably, that due to high kinetic energy and long lifetime of excited $NO_2$ the emission occurs outside of the apparatus' field of view. The lifetimes of $^2B_2$ and $^2B_1$ states of $NO_2$ are 30 μs and 115 μs, respectively [26], [27]. If we assume that the velocity of the produced $NO_2$ is the same as the initial velocity of the nitromethane molecule before electron impact (i.e. 333.3 m/s), then it will travel approximately 1 cm in 30 μs. Since the field of view of our experimental system is only about 3 mm it is probable that we are not be able to detect $NO_2$ continuum.

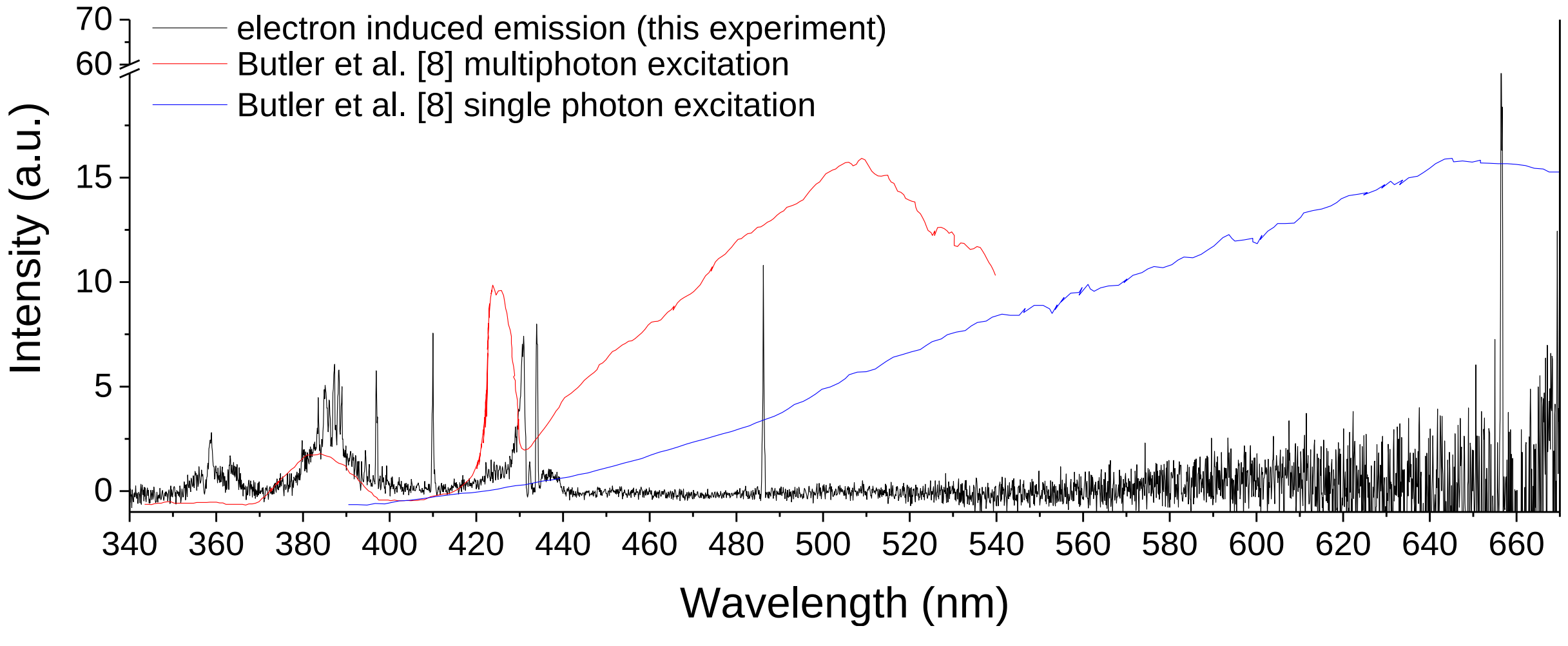

**Figure 7. $NO_2$ continuum radiation from electron impact on CH3NO (50 eV; black) in comparison to laser-induced excitation multiphoton excitation (red); single photon excitation (blue). The continuum emission from the photon-induced experiments dominates the emission above 440 nm but is not present in the electron impact experiments [8]**

## 3.2 Dissociative excitation process

It is important to note that the measured relative emission cross section curves describe the process of electron-induced dissociative excitation, which is strongly dependent on the electron impact energy. By collecting photons from a narrow spectral window centered on individual transitions, it is possible to measure energy dependence of specific transition if the optical resolution of the system is high enough to ensure there is no overlap with the emission features of other species. In this experiment we only measured the emission cross sections of NH (A), CN (B) and H (n = 4) (Table 4). To obtain sufficient SNR in a reasonable amount of time (limited by the lifetime of the electron gun's filament), we increased the electron current by using a relatively low electron energy resolution of 0.8 eV. Therefore, the uncertainty of the determined threshold values exceeds ± 0.34 eV (1σ). The theoretical threshold energy values were calculated based on the enthalpies of the formation of individual products [28]. Comparing these calculated thresholds with the experimental thresholds allows us to discuss the dissociation channels leading to formation of the excited fragments. The experimental thresholds were determined by fitting the segment of the emission cross section curve around the threshold by a convolution of two functions, one being constant corresponding to background signal at energies below the threshold energy value and the second a power function corresponding to the increase of the measured signal just above the threshold. So, the final fitting function was in the form $B + A * (E_e – T)^C$, where B is the background level, $E_e$ is electron energy, T is the threshold electron energy, and A and C are constants.

**Table 4. Theoretically calculated appearance energies corresponding to selected dissociation channels compared to experimentally determined values. The uncertainties in the experimental thresholds is 0.34 eV (1σ).**

| *Emitting species and wavelength [nm]* | *Dissociation channel* | *Channel nr.* | *Calc. threshold [eV]* | *Exp. threshold [eV]* |
|---|---|---|---|---|
| *NH (A) 336* | e + $CH_3NO_2$ → NH (A) + $CH_2$ + $O_2$ + e | (1) | 12.00 | 11.6 ± 0.34 |
| | e + $CH_3NO_2$ → NH (A) + $CH_2$ + O + O + e | (2) | 17.13 | 20.7 ± 0.34 |
| *CH (B) 389* | e + $CH_3NO_2$ → CH (B) + $H_2$ + $NO_2$ + e | (3) | 10.88 | 11.0 ± 0.34 |
| *H (n = 4) 486.2* | e + $CH_3NO_2$ → H (n = 4) + $CH_2NO_2$ + e | (4) | 17.28 | |
| | e + $CH_3NO_2$ → H (n = 4) + $CH_2$ + $NO_2$ + e | (5) | 20.08 | 21.9 ± 0.34 |
| | e + $CH_3NO_2$ → H (n = 4) + CH + H + $NO_2$ + e | (6) | 24.43 | 27.6 ± 0.34 |

$H_\alpha$ is the most intense line of the Balmer series in the spectrum (Figure 2), but it is close to the end of the detector's spectral sensitivity range, resulting in a lower the signal-to-noise ratio. The cross section of $H_\beta$ at 486.2 nm (Figure 8) was chosen to be examined as a representative of the atomic lines in the spectrum. Its experimental threshold of 21.9 ± 0.34 eV is higher than the theoretical threshold of 20.08 eV, indicating that dissociated fragments have high kinetic energy and/or that the energy is being transferred into vibrational and rotational excitation of dissociation fragments. The experimental threshold energy suggests that the proposed channel (5), leading to formation of an excited hydrogen atom along with $CH_2$ and $NO_2$, is preferred over channel (4). Based on their photo-induced fragmentation experiments, Butler et al. [8] suggested that $CH_3$ is the only primary carbon containing photofragment of nitromethane. This is in good agreement with photofragmentation measurements of Blais [29], who suggested that most of the excess energy of the $CH_3$ fragment remains as the internal energy. Moss et al. [30] also considers C-N bond cleavage and production of $CH_3$ to be primary photodissociation processes. Hence, the H (n = 4) is probably produced by the dissociation of the $CH_3$ fragment. The difference in experimental and theoretical thresholds for the channel (6) (opening at 27.6 ± 0.34 eV) producing H (n=4) is even larger, suggesting again high kinetic/internal energy of the fragments.

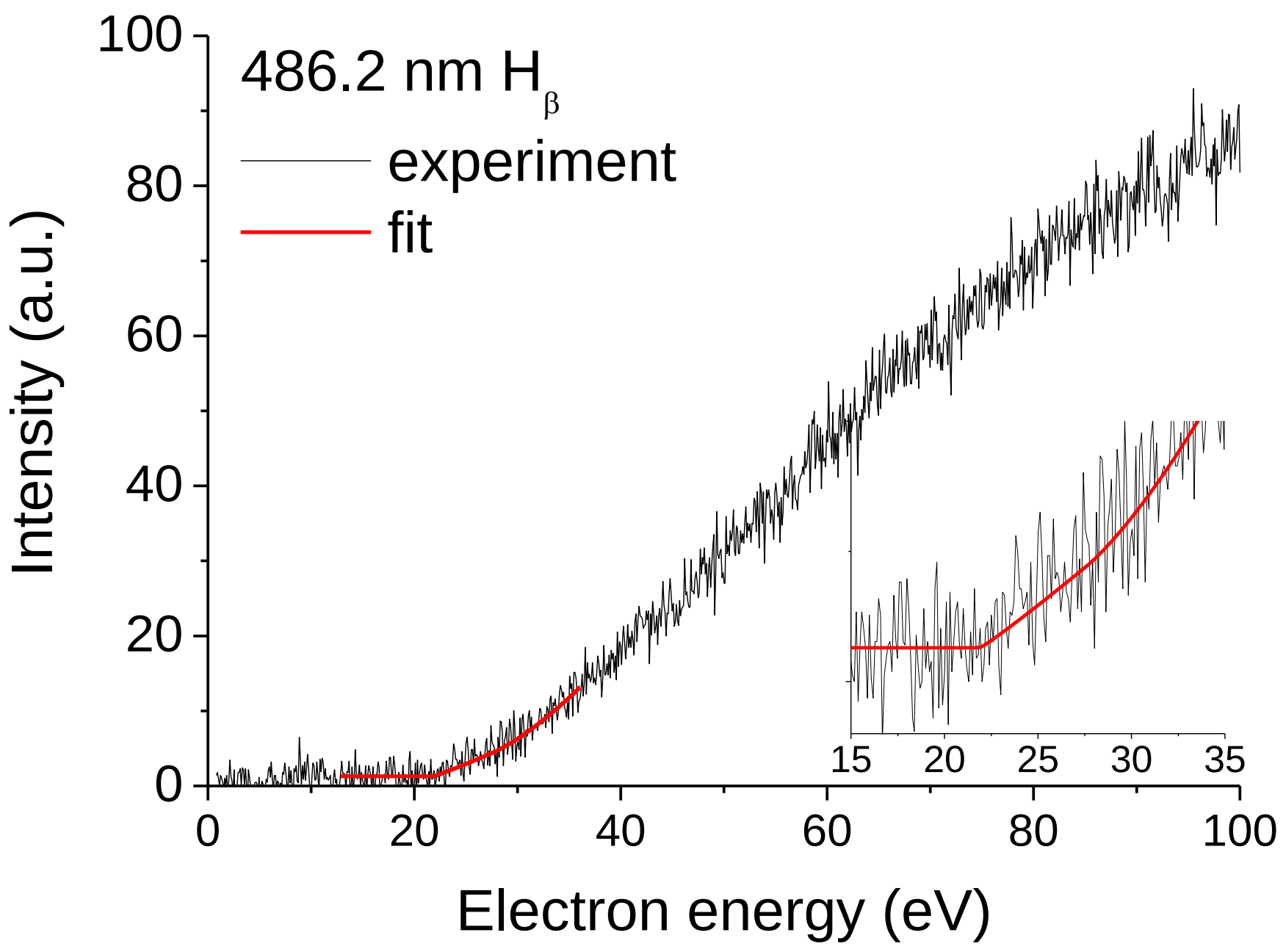

**Figure 8. Relative electron energy-dependent emission cross section of $H_\beta$ Balmer line with the threshold area fit shown in red.**

The emission cross section of the CH (B $^2\Sigma^- \rightarrow$ X $^2\Pi$) (0, 0) feature at 389 nm is shown in Figure 9. The emission of CH (B $^2\Sigma^- \rightarrow$ X $^2\Pi$) (0, 0) (388.9 nm) can be slightly affected by small overlap with CN (B $^2\Sigma^+ \rightarrow$X $^2\Pi$) (0, 0) which is only 0.6 nm away at 388.3 nm. Considering the uncertainties, the measured threshold at 11 +/- 0.34 eV is not higher than theoretical thresholds for both CN (B) (10.77 eV) and CH (B) 10.88 eV. The low energy resolution of the electron gun used in our experiment does not allow us to distinguish between the thresholds for these two processes as our calculations indicate only 0.11 eV difference.

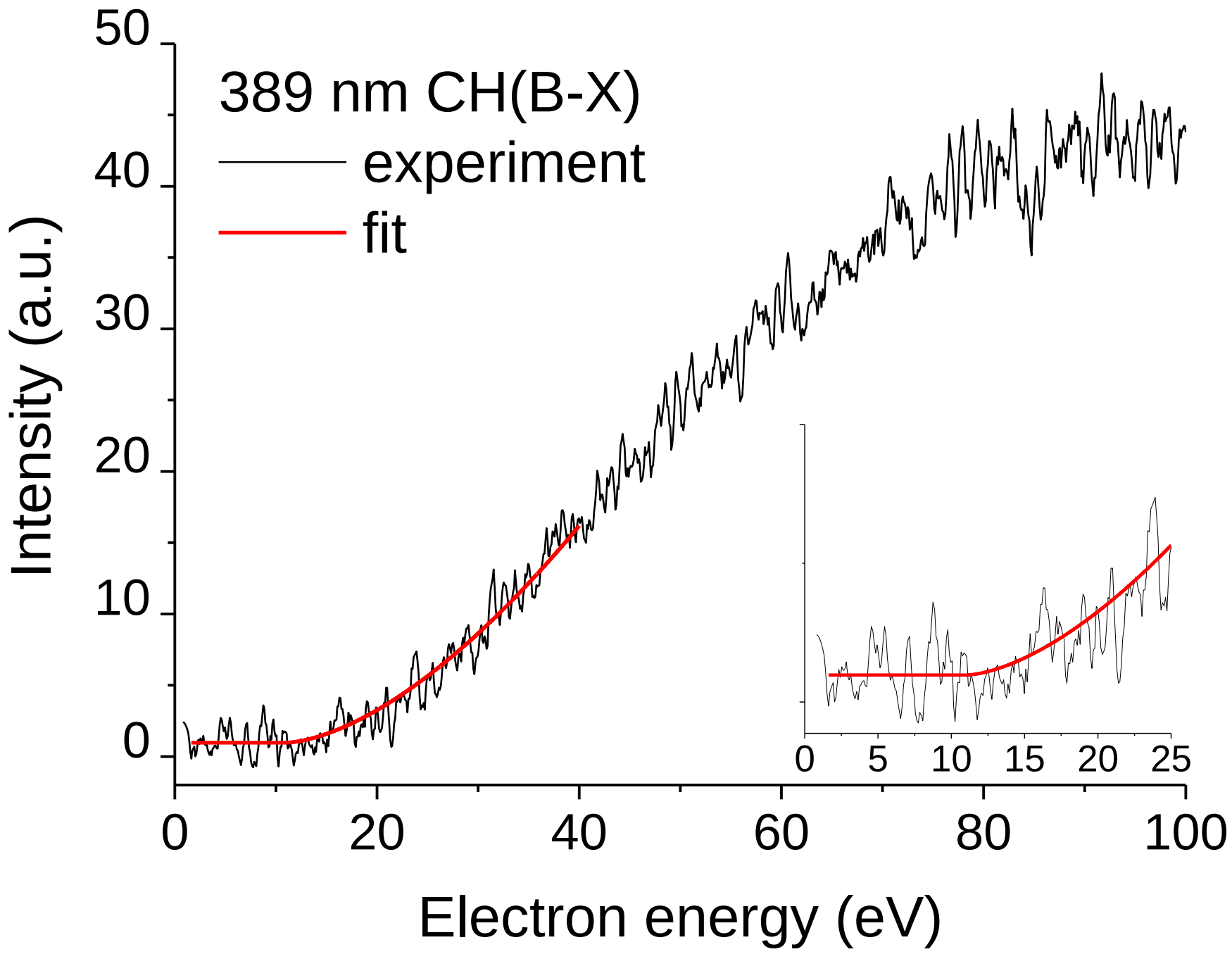

**Figure 9. Relative electron energy-dependent emission cross section of the combined CN (B $^2\Sigma^+ \rightarrow$ X $^2\Pi$) (0, 0) emission at 389 nm.**

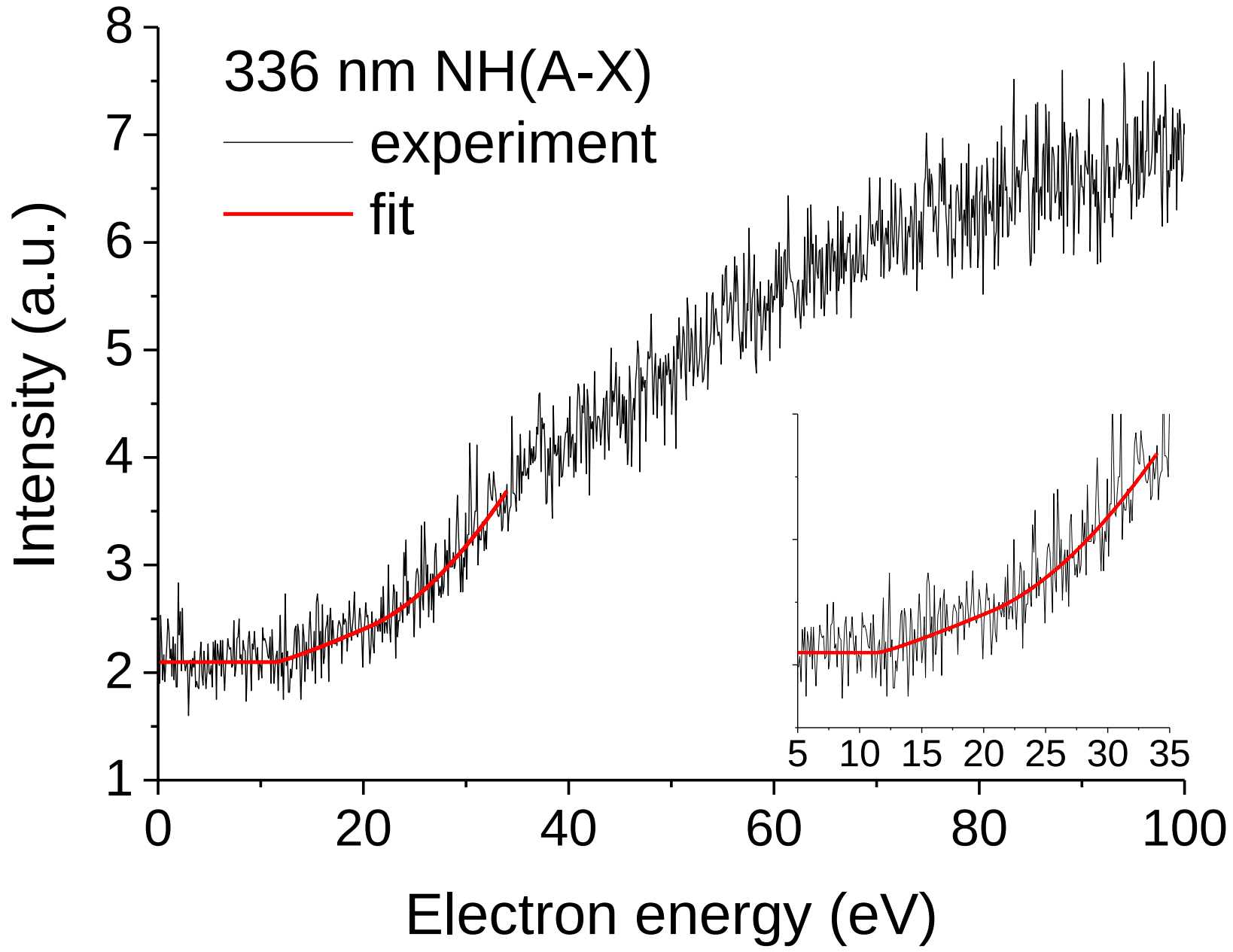

**Figure 10. Relative electron energy-dependent emission cross section of NH (A $^3\Pi \rightarrow X\ ^3\Sigma^-$) (0, 0).**

The emission feature at 336 nm corresponds to the transition of NH (A $^3\Pi \rightarrow X\ ^3\Sigma^-$). In Figure 10, the energy dependence of the cross section of this emission is shown. Since the intensity of the band is lower than the CH or CN transitions at 389 nm, the curve exhibits higher noise and the threshold determination is more complicated and less precise. The first onset of NH (A $^3\Pi$) production via channel (1) at approximately 11.6 ± 0.34 eV is very shallow almost blended in the noise. The curve rise is much steeper at energies above the second threshold at 20.7 ± 0.34 eV suggesting higher efficiency of channel (2). The difference between the calculated and second, experimental threshold value is approximately 3.5 +/- 0.34 eV. The proposed channels (1) and (2) are shown in the Table 4.

# 4 Conclusions

The electron-induced emission spectrum of nitromethane in the range between 200 – 670 nm was recorded at a much higher optical resolution than was previously done. This allowed us to resolve and identify the emission of many atomic and molecular fragments. The most prominent lines and bands in the spectrum corresponded to hydrogen Balmer series, CH, CN violet, NH violet and NO β and γ systems. Although very low in intensity, our results indicate OH radical formation. High threshold energies of the cross section of two overlapping bands, CN (B $^2\Sigma^+ \rightarrow X\ ^2\Pi$) and CH (B $^2\Sigma^- \rightarrow X\ ^2\Pi$) around 389 nm, indicates efficient energy transfer to internal degrees of freedom of nitromethane fragments. The presence of CN tail bands in emission spectrum obtained at 50 eV supports this conclusion. Unlike in photon-induced experiments, we have not detected the $NO_2$ continuum radiation at wavelengths above 400 nm which we attribute to the long lifetimes of the $NO_2$ excited states.

### Acknowledgments

This research was partially supported by Slovak Research and Development Agency under project nr. APVV-19-0386 and by Grant agency VEGA under project Nr. 1/0489/21. This project has received funding from the European Union's Horizon 2020 research and innovation programme under grant agreement No 871149.